\documentclass[a4paper,fleqn,usenatbib]{mnras}

\usepackage{newtxtext,newtxmath}

\usepackage[T1]{fontenc}
\usepackage{ae,aecompl}

\usepackage{graphicx}	
\usepackage{amsmath}	
\usepackage{amssymb}	
\newcommand{\tnrv}[1]{{\textcolor{black}{#1}}}

\title[Red-spiral kinematics]{New constraints on red-spiral galaxies from their kinematics in clusters of galaxies}

\author[A. Hamabata et al.]{
Akinari Hamabata,$^{1}$\thanks{E-mail:hamabata@utap.phys.s.u-tokyo.ac.jp}
Taira Oogi,$^{2}$
Masamune Oguri,$^{1,2,3}$
\newauthor
\   Takahiro Nishimichi$^{4}$
and Masahiro Nagashima$^{5}$
\\
$^{1}$Department of Physics, The University of Tokyo, 7-3-1 Hongo, Bunkyo-ku, Tokyo 113-0033, Japan\\
$^{2}$Kavli Institute for the Physics and Mathematics of the Universe, University of Tokyo, Kashiwa, Chiba 277-8583, Japan\\
$^{3}$Research Center for the Early Universe, The University of Tokyo, 7-3-1 Hongo, Bunkyo-ku, Tokyo 113-0033, Japan\\
$^{4}$Center for Gravitational Physics, Yukawa Institute for Theoretical Physics, Kyoto University, Kyoto 606-8502, Japan\\
$^{5}$Faculty of Education, Bunkyo University, 3337 Minami-Ogishima, Koshigaya-shi, Saitama 343-8511, Japan
}

\date{Accepted XXX. Received YYY; in original form ZZZ}

\pubyear{2015}
\begin{document}
\label{firstpage}
\pagerange{\pageref{firstpage}--\pageref{lastpage}}
\maketitle

\begin{abstract}

The distributions of the pairwise line-of-sight velocity between galaxies and their host clusters are segregated according to the galaxy's colour and morphology.
We investigate the velocity distribution of red-spiral galaxies, which represents a rare population within galaxy clusters.
We find that the probability distribution function of the pairwise line-of-sight velocity $v_{\rm{los}}$ between red-spiral galaxies and galaxy clusters has a dip at $v_{\rm{los}} = 0$, which is a very odd feature, at 93\% confidence level.
To understand its origin, we construct a model of the phase space distribution of galaxies surrounding galaxy clusters in three-dimensional space by using cosmological $N$-body simulations.
We adopt a two component model that consists of the infall component, which corresponds to galaxies that are now falling into galaxy clusters, and the splashback component, which corresponds to galaxies that are on their first (or more) orbit after falling into galaxy clusters.
We find that we can reproduce the distribution of the line-of-sight velocity of red-spiral galaxies with the dip with a very simple assumption that red-spiral galaxies
reside predominantly in the infall component, regardless of the choice of the functional form of their spatial distribution.
Our results constrain the quenching timescale of red-spiral galaxies to a few Gyrs, and \tnrv{the radius where the morphological transformation is effective as $r \sim 0.2 h^{-1} \rm{Mpc}$.}
\end{abstract}



\section{Introduction}
The galaxy evolution is one of the most important topics in modern astrophysics.
Although galaxies have been observed in various ways and are known to have many features such as colours, stellar masses, star formation rate densities, and morphologies, the unified theory to describe the relation between all these features and its evolution has not yet been established.
The environment in which galaxies reside is one of the key factors that determines the features of galaxies.
For instance, it has been known that in the denser region (i.e., cluster-like environments), red, passive, and early-type galaxies are dominant, whereas in the less dense region (i.e., field-like environments), blue, star-forming, and late-type galaxies are dominant (e.g., \citealt{Dressler1980}; \citealt{Peng2010}).

There are several mechanisms that drive such environmental dependence, including the effects of the interactions with other galaxies (e.g., \citealt{Toomre1972}, \citealt{Okamoto2001}), ram pressure stripping due to the pressure exerted by the intra-cluster medium \citep{Gunn1972}, and the strangulation effect, which is the lack of the gas in galaxies due to the cutoff of the gas stock (\citealt{Larson1980}; \citealt{Balogh2000}).
These mechanisms can quench star formation of galaxies and turn them into red, passive, and early-type galaxies.
These mechanisms are implemented to cosmological simulations, and the relation between the quenching mechanism and the resulting distribution of galaxies is also investigated (e.g., \citealt{Gabor2012}; \citealt{Lotz2018}).

It has been known that the majority of galaxies follow the tight colour-morphology relationship.
\citet{Lintott2008} showed that red galaxies are dominated by early-type galaxies, and the majority of blue galaxies are late-type galaxies at $z< 0.12$.
This tight relationship can be explained as follows.
The red galaxies tend to have old stellar populations, and the early-type galaxies are the end point of the dynamical history of galaxies.
In addition, timescales of mechanisms that drive the morphological transformation and quench star formation are strongly related in most cases.

Red-spiral galaxies are the rare subpopulation of galaxies, whose colours are red, and the morphology is late-type, i.e., old stellar-population and young dynamical state.
The red-spiral galaxies are investigated in some previous papers.
For example, \citet{Masters2010} studied spectroscopic properties and environments of passive red-spiral galaxies found in the Galaxy Zoo project \citep{Lintott2008}.
One of the possible scenarios which explains the red-spiral galaxies is that red-spiral galaxies are accreted on to massive haloes as blue-spirals, after which their outer halo gas reservoirs are stripped by environmental effects without transforming their morphology, and quench the star formation (e.g., \citealt{Bekki2002}).
Red-spiral galaxies then turn into another morphological type of galaxies by merging with other galaxies and/or their spontaneous dynamical evolution.
This scenario is consistent with the observational results shown in \citet{Masters2010}, although other reasonable scenarios that explain the red-spiral galaxies has also been proposed (e.g., \citealt{Masters2010}; \citealt{Schawinski2014} ).
The identification of the detailed scenario for the formation of
red-spiral galaxies provides an important clue for understanding the origin of the colour-morphology relationship of galaxies, and therefore crucial for obtaining the whole picture of galaxy evolution.
This is why new observational clues to constrain the scenario are anticipated.

The distribution of the pairwise line-of-sight velocity ($v_{\rm{los}}$) between galaxies and their host clusters has been used to estimate the depths of the gravitational potential and to infer the masses of their host clusters (e.g., \citealt{Smith1936}; \citealt{Busha+2005}).
Furthermore, the segregations of the distribution of $v_{\rm{los}}$ by galaxy types have also been studied.
Some studies have found that the dispersion of the probability distribution function (PDF) of $v_{\rm{los}}$ for blue, late-type, and star-forming galaxies tends to be larger than that for red, early-type, and passive galaxies (e.g., \citealt{Sodre1989}; \citealt{Bivianoo2002}; \citealt{Bayliss2017}).
It has also been found that the PDF of $v_{\rm{los}}$ for fainter galaxies has the larger dispersion than that for luminous galaxies (e.g., \citealt{Chincarini1977}; \citealt{Goto2005}).
Since the segregations of the distribution of $v_{\rm{los}}$ between galaxies and their host clusters
contain a lot of information about the properties of galaxy populations in the most dense environments, it is important to use the full distribution of line-of-sight velocities, not just the dispersion of the PDF as often studied in the literature.
Indeed, in some studies (e.g., \citealt{Oman2016}, \citealt{Rhee2017}, \citealt{Adhikari2018} \citealt{Arthur2019}), the segregation of the full distribution of $v_{\rm{los}}$ has been studied using galaxies obtained from cosmological simulations.
However, a proper interpretation of the observed phase-space data is not been straightforward because it requires the knowledge of the full phase space distribution and taking a proper account of projection along the line-of-sight.

Red-spiral galaxies are a very rare class of galaxies.
In addition, it is difficult to obtain the morphological information of galaxies from observations. 
Hence, the PDF of $v_{\rm{los}}$ between red-spiral galaxies and their host clusters has not been measured because of the large statistical error originating from a small number of red-spiral galaxies.
With the help of the morphological information for a large number of galaxies from Galaxy Zoo project \citep{Lintott2008} and galaxy cluster catalogues that contain a large number of galaxy clusters (e.g., \citealt{Rykoff2014}; \citealt{Oguri2014}), in this paper we investigate the PDF of $v_{\rm{los}}$ between red-spiral galaxies and galaxy clusters by stacking different clusters to reduce the statistical error.

\citet{Hamabata2018} studied the relationship between the PDF of $v_{\rm{los}}$ between galaxies and galaxy clusters, the phase space distribution around clusters in three-dimensional space, and the radial distribution of galaxies by using cosmological $N$-body simulations.
In this paper, we adopt the model of \citet{Hamabata2018} to interpret the observed PDF of $v_{\rm los}$ for red-spiral galaxies, by taking proper account of the observed radial distribution of red-spiral galaxies.
Our approach allows us to make full use of the phase space distribution information obtained from the observation, not just the dispersion of the PDF, and extract the information that the observed PDF contains properly.

This paper is organised as follows.
In Section \ref{S-Obs}, we show the observational data and our analysis method.
In Section \ref{S-Sim}, we present our model of the phase space distribution of galaxies in three-dimensional space from cosmological $N$-body simulations.
In Section \ref{S-Comp}, we compare the results obtained from the observation and our simulations.
In Section \ref{S-Dis}, we provide a detailed discussion of the results.
Finally, we conclude in Section \ref{S-Con}.
We adopt cosmological parameters $h = 0.7$, $\Omega_{M,0} = 0.279$, $\Omega_{\Lambda,0} = 0.721$, $n_{s} = 0.972$, and $\sigma_{8} = 0.821$ following the WMAP nine year result \citep{Hinshaw2013}  throughout this paper.

\section{Result From Observational Data}
\label{S-Obs}
\subsection{Data of Clusters and Galaxies}
To determine the kinematics of cluster galaxies, we require precise spectroscopic redshifts.
For this purpose, we employ the spectroscopic galaxy sample from SDSS DR10 data \citep{DR10}
We use galaxies observed by Legacy spectroscopic observation \citep{SDSS} only, because we have to use galaxies selected by uniform criteria.
Following \citet{Masters2010}, we select red galaxies based on their $\tt{cModelMag}$, as $ (g - r) > 0.63 - 0.02   ( M_{r} + 20)$, where $M_{r}$ is the absolute magnitude in $r$ band.
We also limit the galaxy sample for our analysis to $M_{r} < -20.17$.
For each galaxy, the absolute magnitude is computed from its apparent {\tt cModelMag} with the k-correction using the technique defined in \citet{Chilingarian2012}, and also with the correction of Galactic extinction \citep{Schlegel1998}.
\begin{figure}
    \includegraphics[width=\columnwidth]{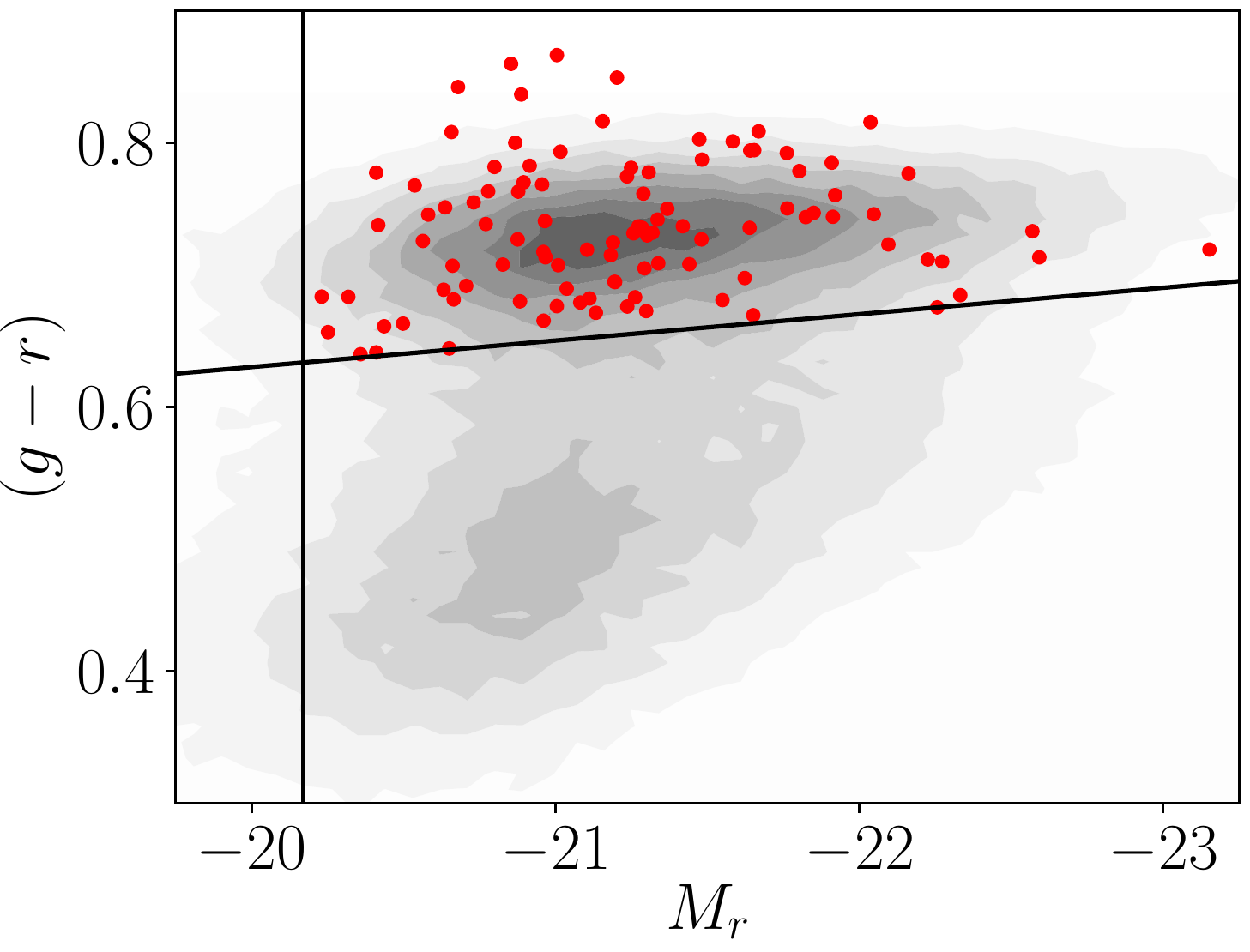}
 \caption{Colour-magnitude diagram of our galaxy sample.
Red points indicate red-spiral galaxies at $ | z_{\rm{g}} - z_{\rm{c}} |< 0.01$ within the transverse distance of $ 0.5  ~ h^{-1} {\rm{Mpc}} $ from the centre of galaxy clusters at $0.05<z<0.1$ that we use in this analysis.
The contours show the number distribution of the Galaxy Zoo 1 sample in the redshift range $ 0.05<  z < 0.1$.
The solid lines indicate the colour cut and the absolute magnitude cut adopted in this paper.}
 \label{F-Cont}
\end{figure}

Galaxy Zoo is an online citizen science project to obtain morphological \tnrv{information} for a large number of galaxies from visual inspections.
We use the Galaxy Zoo 1 data release \citep{Lintott2008}, which is the largest morphologically classified sample of galaxies.
We select spiral galaxies as $p_{\rm{cs \_ debiased}} > 0.8$, where $p_{\rm{cs \_ debiased}}$ is the debiased fraction of votes for combined spiral galaxies with the correction of the selection bias and the classification bias of the Galaxy Zoo sample in \citet{Bamford2009}.
\tnrv{We can only obtain $p_{\rm{cs \_ debiased}}$ and $p_{\rm{el \_ debiased}}$ as  debiased morphological information of galaxies, where $p_{\rm{el \_ debiased}}$ is the debiased fraction of votes for elliptical galaxies with the corrections.
Hence, we cannot extract information of other morphological types of galaxies, such as S0 galaxies.}

In this paper, we use a cluster sample constructed by a red-sequence cluster method, in which clusters are selected by overdensities of red galaxies (e.g., \citealt{Rykoff2014}; \citealt{Oguri2014}).
Here we adopt the SDSS DR8 redMaPPer cluster catalogue \citep{Rozo2015}, because the redMaPPer cluster catalogue contains clusters down to the sufficiently low redshift of $z=0.05$ and hence has a large overlap of the redshift range with the Galaxy Zoo galaxy sample.

In Fig. \ref{F-Cont}, we show the colour-magnitude diagram of red-spiral galaxies that we use in this paper.
For comparison, we also show the distribution of the Galaxy Zoo 1 sample.
Our sample consists of 968 red galaxies and 101 red-spiral galaxies at $ | z_{\rm{g}} - z_{\rm{c}} |< 0.01$ within the transverse distance of $ 0.5  ~ h^{-1} {\rm{Mpc}} $ from the centre of galaxy clusters at $0.05<z<0.1$ that we use in this analysis, where $z_{\rm{g}}$ and $z_{\rm{c}}$ are redshifts of the galaxy and the cluster, respectively.

\subsection{The Observed PDF of the line-of-aight velocity $v_{\rm{los}}$}
\label{S-ObsA}
The line-of-sight velocity $v_{\rm{los}}$ between a galaxy and a cluster is given by
\begin{equation}
 \label{ObsA-1}
  v_{\rm{los}} = c \  \left ( \frac{z_{\rm{g}} - z_{\rm{c}}}{1 + z_{\rm{c}}} \right ).
\end{equation}
Throughout the paper we adopt the spectroscopic redshift of the central galaxy as the redshift of each cluster.
Because we are interested in average features of the kinematics of galaxies, we stack galaxies around clusters to obtain statistically better sample.
We apply a richness cut $20 < \lambda < 40$,  where $\lambda$ is the richness of each cluster estimated in the redMaPPer.
We also adopt a redshift cut $0.05 < z_{c} < 0.1$, and only stack clusters whose central galaxies have spectroscopic redshifts.
Finally, we stack 90 clusters in this work.
\begin{figure}
    \includegraphics[width=\columnwidth]{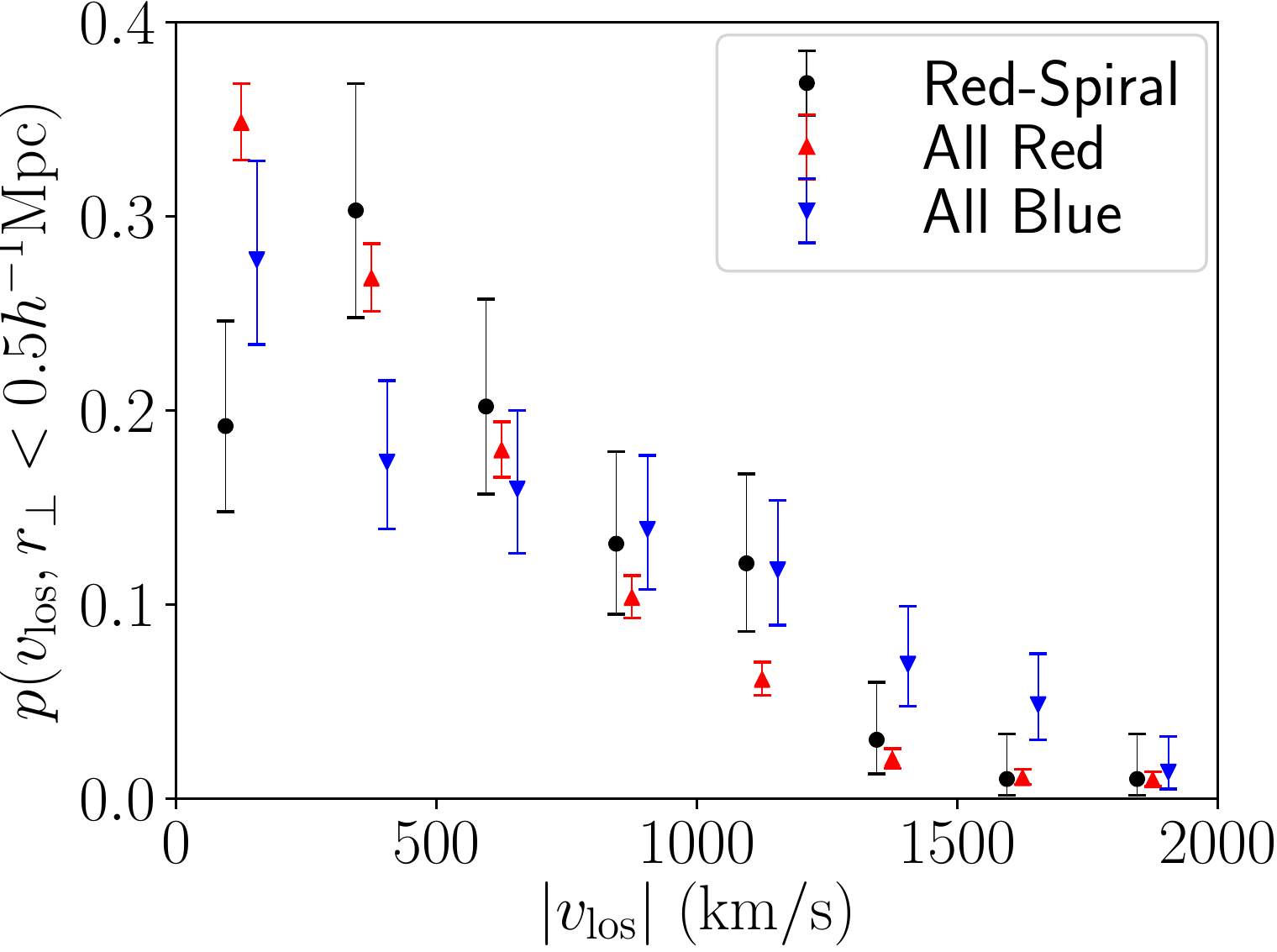}
 \caption{\tnrv{The PDF of $|v_{\rm{los}}|$ of red-spiral galaxies ($\it{black ~ circle}$), red galaxies ($\it{red ~ triangle ~ up})$, and blue galaxies ($\it{blue ~ triangle ~ down}$).}
Galaxies within projected $0.5 h^{-1} \rm{Mpc}$ from the centre of the clusters are used.}
 \label{F-ObsA-1}
\end{figure}

In Fig. \ref{F-ObsA-1}, we show the PDF of the line-of-sight velocity, $p (|v_{\rm{los}}|)$, for red-spiral galaxies.
For comparison, we also show $p (|v_{\rm{los}}|)$ for all red galaxies, which are galaxies with $ (g - r) > 0.63 - 0.02  ( M_{r} + 20)$, $M_{r} < -20.17$, and $p_{\rm{cs \_ debiased}} \leq  1.0$, \tnrv{and all blue galaxies, which are galaxies with $ (g - r) < 0.63 - 0.02  ( M_{r} + 20)$, $M_{r} < -20.17$, and $p_{\rm{cs \_ debiased}} \leq  1.0$}.
We stack galaxies within projected $0.5 h^{-1} \rm{Mpc}$ from the centres of the clusters.
The error bars are Poisson errors from the number of galaxies in each $|v_{\rm{los}}|$ bin.
We can see that the PDF of $v_{\rm{los}}$ for red-spiral galaxies has a dip at $v_{\rm los}=0$, which is not seen in the PDF for all red galaxies \tnrv{nor all blue galaxies.}
The statistical significance of the dip as measured by the difference of the first two bins is $1.4 \sigma$ ($93 \%$ C.L.).
By using the Kolmogorov-Smirnov test, we find that $p (|v_{\rm{los}}|)$ for red-spiral galaxies and that for all red galaxies are different from each other at the significance higher than 99.99\%.

\section{Phase Space Distribution in three-dimensional space from Simulations}
\label{S-Sim}

To interpret the observational data, we construct a model of the distributions of $v_{\rm{los}}$ between the galaxy and the cluster.
The PDF of $v_{\rm{los}}$ can be described as
 \begin{equation}
 \label{eqS-01}
 \begin{split}
& p_{v_{\rm{los}}} (v_{\rm{los}} , r_{\perp})  = \\
 & \frac{1}{N (r_{\perp})} \iiint d \vec{v}  \int d r_{\parallel}   \rho ( \vec{r})  p_{v}( \vec{v} , \vec{r})  \delta_{D} ( v_{\rm{los}} - v'_{\rm{los}} ) ,
\end{split}
\end{equation}
where $\vec{v}$ is the three-dimensional peculiar pairwise velocity between the galaxy and the cluster, $\vec{r}$ is the physical spatial coordinates of the galaxy relative to the cluster, $r_{\perp}$ is the transverse distance from the cluster centre, $r_{\parallel}$ is the line-of-sight distance from the cluster centre, $\rho (\vec{r})$ is the spatial distribution of galaxies, $\delta_{D} (x)$ is the Dirac's delta function, $p_{v}( \vec{v} , \vec{r})$ is phase space distribution in three-dimensional space, and $N (r_{\perp})$ is the normalisation factor.
Here, $v'_{\rm{los}}$ is defined as
 \begin{equation}
 \label{eqS-02}
v'_{\rm{los}} \equiv v_{\parallel} + \frac{H(z)  r_{\parallel}}{1 + z} ,
\end{equation}
where $v_{\parallel}$ is the parallel component to the line-of-sight of the peculiar pairwise velocity between the galaxy and the cluster, and $H(z)$ is the Hubble parameter.
As shown in equation (\ref{eqS-01}), to construct a model of the distributions of $v_{\rm{los}}$,  we have to construct a model of the phase space distribution in three-dimensional space.
In this Section, we present our model of the phase space distribution in three-dimensional space, based on the model presented in Section 2 of \citet{Hamabata2018}.

\subsection{Simulations}
We use four random realisations of cosmological $N$-body simulations with a TreePM code $\tt{Gadget}$-$2$ \citep{Springel2005}.
These simulations run from $z = 99$ to $0$, the box size is comoving $360 ~ h^{-1} \rm{Mpc} $ on a side, and the gravitational softening length is comoving $20  ~ h^{-1} \rm{kpc} $.
They are performed with the periodic boundary condition.
The number of dark matter particles is $1024^{3}$ with $m_{p} = 3.4 \times 10^{9}  h^{-1} M_{\odot} $.
In setting up the initial conditions, we first compute the linear matter power spectrum at $z=0$ using a Boltzmann solver $\tt{CAMB}$ \citep{Lewis2000}.
We then scale it to $z=99$ by the linear growth factor computed for the $\Lambda$CDM cosmology ignoring radiation or relativistic corrections.
Subsequently, a Gaussian random field for the linear overdensity $\delta_{\rm lin}(z_{\rm in})$ is generated following this power spectrum.
We finally compute the displacement field for the fluid elements located at a regular lattice up to the second order in $\delta_{\rm lin}(z_{\rm in})$ using a code developed in \citet{Nishimichi2009} and parallelised in \citet{Valageas2011}.
We use six-dimensional friend of friend (FoF) algorithm implemented in $\tt{Rockstar}$ \citep{Behroozi2013B}.
While the $\tt{Rockstar}$ identifies both haloes and subhaloes from $N$-body simulations, we do not distinguish them and call both of them haloes.

\subsection{Stacked Phase Space Distribution}
We use haloes with masses $M_{200 m}> 1 \times 10^{11} h^{-1} M_\odot $ to represent galaxy haloes in this work, where $M_{200 m}$ is the mass within $r_{200 m}$, which is the radius within which the average density is 200 times the background matter density at the redshift of interest.
We use haloes with masses $ 9.5 \times 10^{13} h^{-1} M_\odot  <  M_{200 m} < 2.3 \times 10^{14} h^{-1} M_\odot$ to represent cluster haloes, which corresponds to galaxy clusters with $20 < \lambda < 40$ given the mass-richness relation shown in \cite[see also \citealt{Murata2018}]{Simet2018}, where $\lambda$ is the richness of galaxy clusters calculated in the redMaPPer.
To mimic the observed cluster catalogue, we remove cluster haloes from our analysis if there are any other cluster haloes with larger masses within physical $1  ~ h^{-1} \rm{Mpc} $ from those cluster haloes.
We use only snapshots at $z = 0.1$ in this work.

To obtain accurate phase space distributions in three-dimensional space, we stack a large number of pairs of galaxy haloes and cluster haloes from these simulations without aligning their orientations.
The peculiar pairwise velocity between cluster haloes and galaxy haloes can be divided into three orthogonal components, the radial velocity ($v_{r}$) and two tangential velocities ($v_{t;1} , \ v_{t;2} $).
Since we can choose the coordinates such that one tangential velocity component is always orthogonal to the line-of-sight, we can ignore $v_{t;2} $ and denote $v_{t;1}$ as $v_{t}$.
The asphericity of cluster haloes tends to disappear after stacking, and the resultant averaged phase space should be fully specified by the subspace of $(r, v_{r}, v_{t})$.
Indeed, we are not interested in the asphericity of the observed clusters, and a model in this subspace is sufficient to obtain the projected phase space $(r_{\perp}, v_{\rm{los}})$.

We adopt a two component model of the phase space distribution presented in \citet{Hamabata2018}.
This model assumes that the PDF of the phase space distribution can be divided into two components, the infall component (IF) and the splashback component (SB).
The PDF of the phase space distribution can be derived as
 \begin{equation}
 \label{eqS-1}
 \begin{split}
p_{v}(v_{r} , v_{t} ,r) = \ & (1 - \alpha )  p_{\rm{infall}} (v_{r}, v_{t} ,r)  + \alpha   p_{\rm{SB}} (v_{r}, v_{t}  ,r)\ ,
\end{split}
\end{equation}
where $p_{\rm{infall}}$ and $p_{\rm{SB}}$ are properly normalised, and $\alpha$ denotes the fraction of the splashback component at given $r$.
The first term of the right hand side of equation (\ref{eqS-1}) represents the infall component, whose average value of the radial velocity is negative.
Thus, the infall component corresponds to galaxy haloes that are now falling into cluster haloes.
On the other hand, the send term of the right hand side of equation (\ref{eqS-1}) represents the splashback component, whose average value of the radial velocity is positive.
The splashback component which corresponds to galaxy haloes that are on their first (or more) orbit after falling into cluster haloes.
Following \citet{Hamabata2018}, we ignore the correlation between $v_{r}$ and $v_{t}$ for both the two components, and rewrite equation (\ref{eqS-1}) as
 \begin{equation}
 \label{eqS-1-1}
 \begin{split}
p_{v}(v_{r} , v_{t} ,r) = \ & (1 - \alpha )  p_{v_{r},\rm{infall}} (v_{r} ,r)  p_{v_{t},\rm{infall}} (v_{t} ,r) \\
& + \alpha   p_{v_{r},\rm{SB}} (v_{r}  ,r)  p_{v_{t},\rm{SB}} (v_{t}  ,r)\ .
\end{split}
\end{equation}
Again, $p_{v_{r},\rm{infall}}$, $p_{v_{t},\rm{infall}}$, $p_{v_{r},\rm{SB}}$, and $p_{v_{t},\rm{SB}}$ are properly normalised.
While there is a room to improve the two component model, for example by adding more components, we adopt this two component model for simplicity.
We leave the construction of more complex models for future work.
We discuss each component in equation (\ref{eqS-1-1}) in what follows.

\subsubsection{Radial Velocity Distribution}
\begin{figure}
 \begin{minipage}{0.45\textwidth}
    \includegraphics[width=1.0\textwidth]{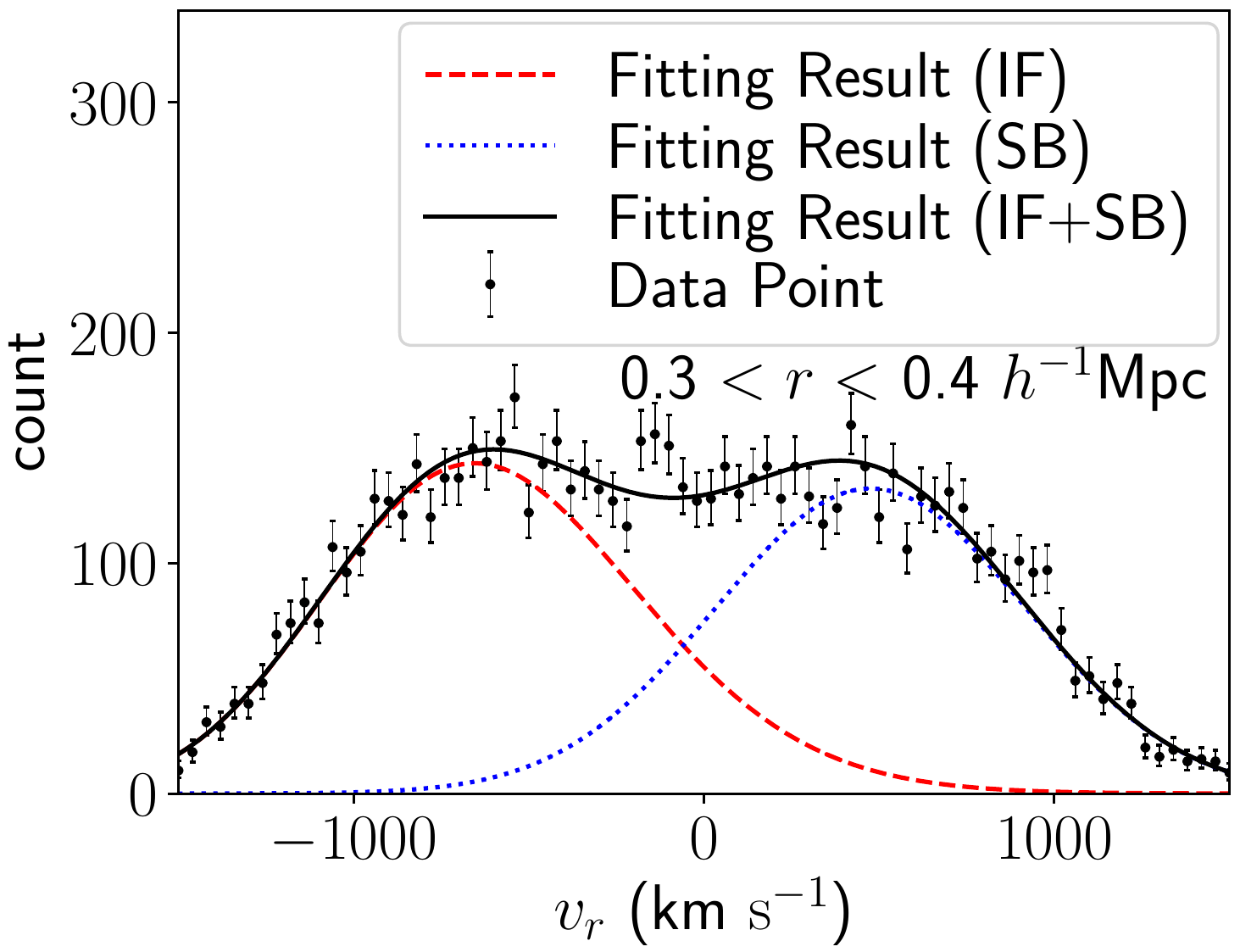}
           \end{minipage}\\
    \begin{minipage}{0.45\textwidth}
    \includegraphics[width=1.0\textwidth]{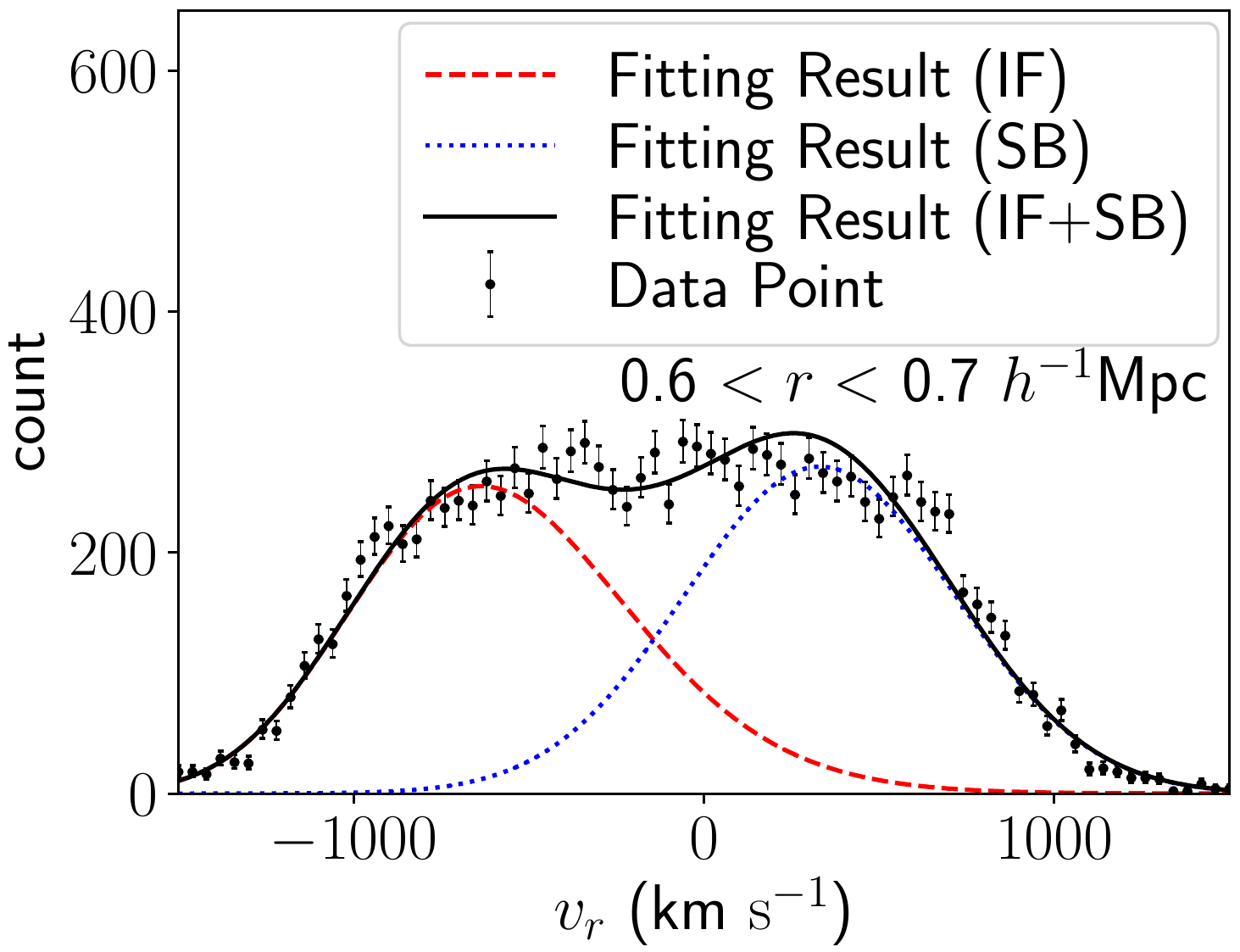}
       \end{minipage}
       \caption{The radial velocity distributions at $0.3  ~ h^{-1} {\rm{Mpc}}  < r < 0.4  ~ h^{-1} {\rm{Mpc}} $ ($\it{top \ pannel}$) and $0.6 ~ h^{-1} {\rm{Mpc}} < r < 0.7  ~ h^{-1} {\rm{Mpc}} $ ($ \it{bottom \ pannel}$).
Points with error bars are the histogram of radial velocities from our $N$-body simulations, the red dashed line is the best fit line of the infall component (IF), the blue dotted line is the best fit line of the splashback component (SB), and the black solid line is the sum of the dashed and dotted lines.}

    \label{FigS-2}
\end{figure}
We start with the distribution of radial velocities.
\citet{Hamabata2018} found that the radial velocity distribution of the infall component have non-negligible skewness and kurtosis, and adopt the Johnson's SU-distribution \citep{Jhonson1949} as the model function for the radial velocity distribution of the infall component.
Following \citet{Hamabata2018}, we adopt the Johnson's SU-distribution as
 \begin{equation}
 \label{eqS-2}
  \begin{split}
p_{v_{r},\rm{infall}} & (v_{r} , r) =\  SU (v_{r};\delta_{r},\lambda_{r},\gamma_{r}, \xi_{r} ) \\
 =\ & \frac{\delta_{r} }{\lambda_{r} \sqrt{2 \pi} \sqrt{ \{ z(v_{r}) \}^2 + 1}} \exp{ \left[ - \frac{1}{2}   \left( \gamma_{r} + \delta_{r} \sinh^{-1}{z(v_{r})} \right )    \right]},
 \end{split}
\end{equation}
where
 \begin{equation}
 \label{eqS-2-3}
z(v_{r}) = \frac{v_{r} - \xi_{r}}{ \lambda_{r}}.
\end{equation}
Note that $\delta_{r},\lambda_{r},\gamma_{r}$, and $\xi_{r}$ are free parameters, and they are functions of the radius $r$.

For the splashback component, we adopt the Gaussian distribution,
 \begin{equation}
 \label{eqS-3}
   \begin{split}
p_{v_{r},\rm{SB}} (v_{r} , r) =\  & G (v_{r} ; \mu_{r}  , \sigma_{r}^{2} ) \\
=\  & \frac{1}{\sqrt{2 \pi \sigma_{r}^2 }} \exp{ \left \{ - \frac{ ( v_{r} - \mu_{r}  ) ^ 2}{ 2 \sigma_{r}^2 } \right \} } \ ,
\end{split}
\end{equation}
where $\mu_{r}$ and $\sigma_{r}$ are free parameters, and they are also functions of $r$.

We constrain these free parameters as follows.
By integrating over the tangential velocity, we can derive the model function of the PDF of the radial velocity distribution from equation (\ref{eqS-2}) as
 \begin{equation}
 \begin{split}
 \label{eqS-4}
p_{v_{r}} (v_{r} , r) =  & \ \int d v_{t}  \{ (1 - \alpha )  p_{v_{r},\rm{infall}} (v_{r} ,r)  p_{v_{t},\rm{infall}} (v_{t} ,r) \\
& + \alpha   p_{v_{r},\rm{SB}} (v_{r}  ,r)  p_{v_{t},\rm{SB}} (v_{t}  ,r) \}  \\
= & \ (1 - \alpha) p_{v_{r},\rm{infall}} + \alpha p_{v_{r},\rm{SB}}\ .
\end{split}
\end{equation}
For a given radial bin, there are seven free parameters in equation (\ref{eqS-4}).
We fit the histogram of the radial velocity obtained from our $N$-body simulations with equation (\ref{eqS-4}) for each radial bins.
We adopt $0.1  ~ h^{-1} {\rm{Mpc}}$ as the width of each radial bin.

Fig. \ref{FigS-2} shows examples of radial velocity distributions and fitting results of equation (\ref{eqS-4}).
Here the error bars are Poisson errors from the number of haloes in each $v_r$ bin.
Our model function of the radial velocity distribution is in good agreement with the histogram from our $N$-body simulations even at $r$ smaller than $2.0  ~ h^{-1} {\rm{Mpc}}$, where \citet{Hamabata2018} did not investigate.
Since the splashback component is negligibly small at large $r$, in fitting we always fix $\alpha = 0$ at $r$ larger than $5  ~ h^{-1} {\rm{Mpc}}$.
This Figure indicates that the absolute value of the average radial velocity of the infall component is larger than that of the splashback component.
By using the fitting results, we interpolate the parameters linearly and use them as smooth functions of $r$.

\subsubsection{Tangential Velocity Distribution}
Next we model the tangential velocities.
We adopt the Johnson's SU-distribution again as a model function for the tangential velocity distribution of both the infall and splashback components.
We adopt the model function of the distributions of tangential velocity as
 \begin{equation}
 \label{eqS-5}
  \begin{split}
p_{v_{t},X} & (v_{t} , r) =  SU (v_{t};\delta_{X:t},\lambda_{X:t} ) \\
 =\ & \frac{\delta_{X:t} }{\lambda_{X:t} \sqrt{2 \pi} \sqrt{ \{ z(v_{t}) \}^2 + 1}} \exp{ \left[ - \frac{\delta_{X:t}}{2}  \sinh^{-1}{z_{X}(v_{t})}  \right]},
 \end{split}
\end{equation}
where
\begin{equation}
 \label{eqS-6}
z_{X}(v_{t}) = \frac{v_{t} }{ \lambda_{X:t}} \ ,
\end{equation}
and $X$ runs over infall and SB, and $\delta_{X:t}$ and $\lambda_{X:t}$ are free parameters as functions of $r$.
The odd-order moments of the PDF of the tangential velocity distribution must be zero, because we assume the spherical symmetry.
Hence, we set the $\gamma_{X:t}$, and $\xi_{X:t}$ in the Johnson's SU-distribution to zero.

The model function for the tangential distribution is
 \begin{equation}
   \begin{split}
 \label{eqS-7}
p_{v_{t}} (v_{t} , r) = & \ \int d v_{r} \{ (1 - \alpha )  p_{v_{r},\rm{infall}} (v_{r} ,r)  p_{v_{t},\rm{infall}} (v_{t} ,r) \\
& + \alpha   p_{v_{r},\rm{SB}} (v_{r}  ,r)  p_{v_{t},\rm{SB}} (v_{t}  ,r) \} \\
= & \ (1 - \alpha)  p_{v_{t},\rm{infall}} + \alpha p_{v_{t},\rm{SB}}\ .
  \end{split}
\end{equation}
There are four free parameters in equation (\ref{eqS-7}) for the single radial bin, because we fix $\alpha$ to the value that is obtained from the fitting of the radial velocity distribution at the same radial bin.

\begin{figure}
 \begin{minipage}{0.45\textwidth}
    \includegraphics[width=1.0\textwidth]{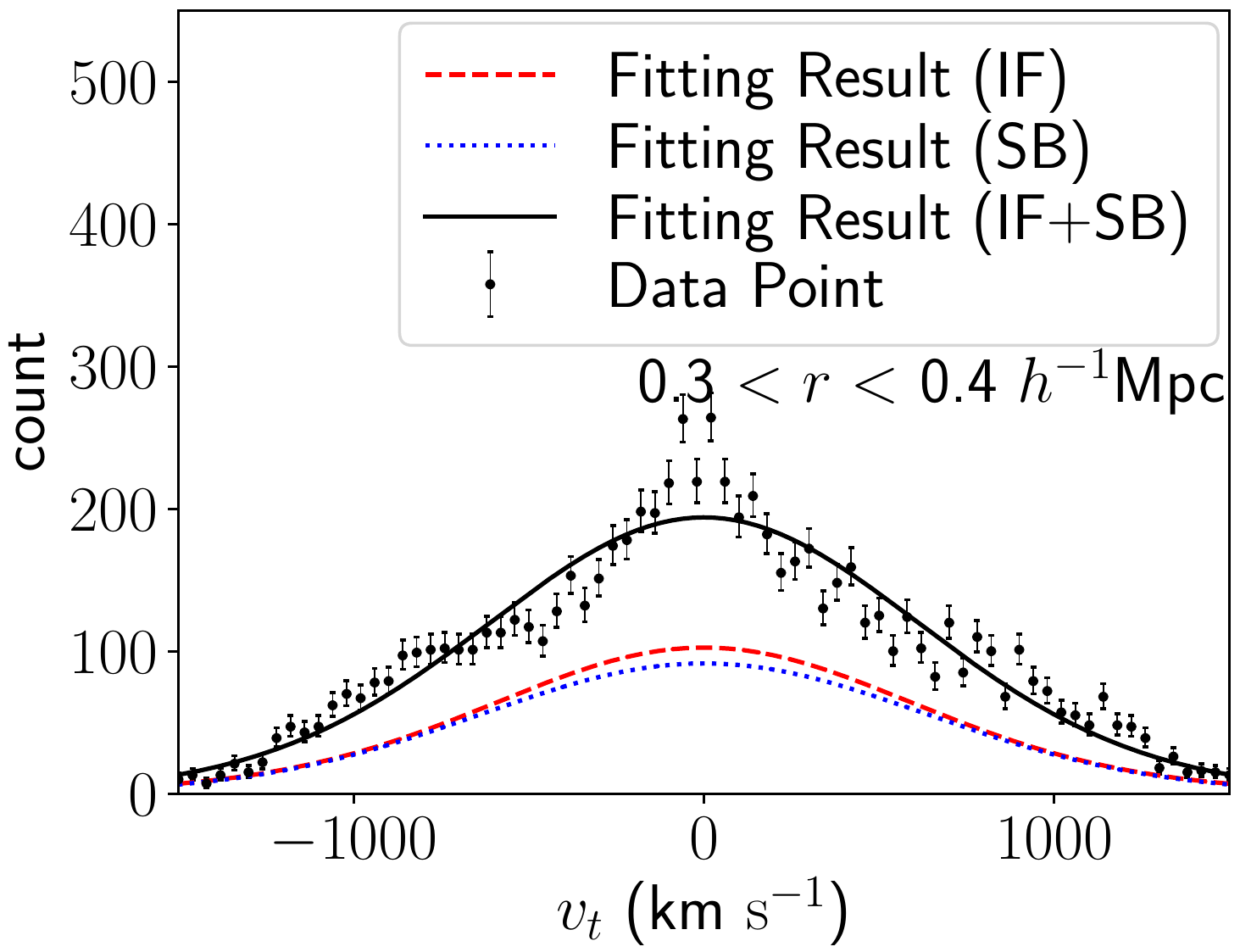}
           \end{minipage}\\
    \begin{minipage}{0.45\textwidth}
    \includegraphics[width=1.0\textwidth]{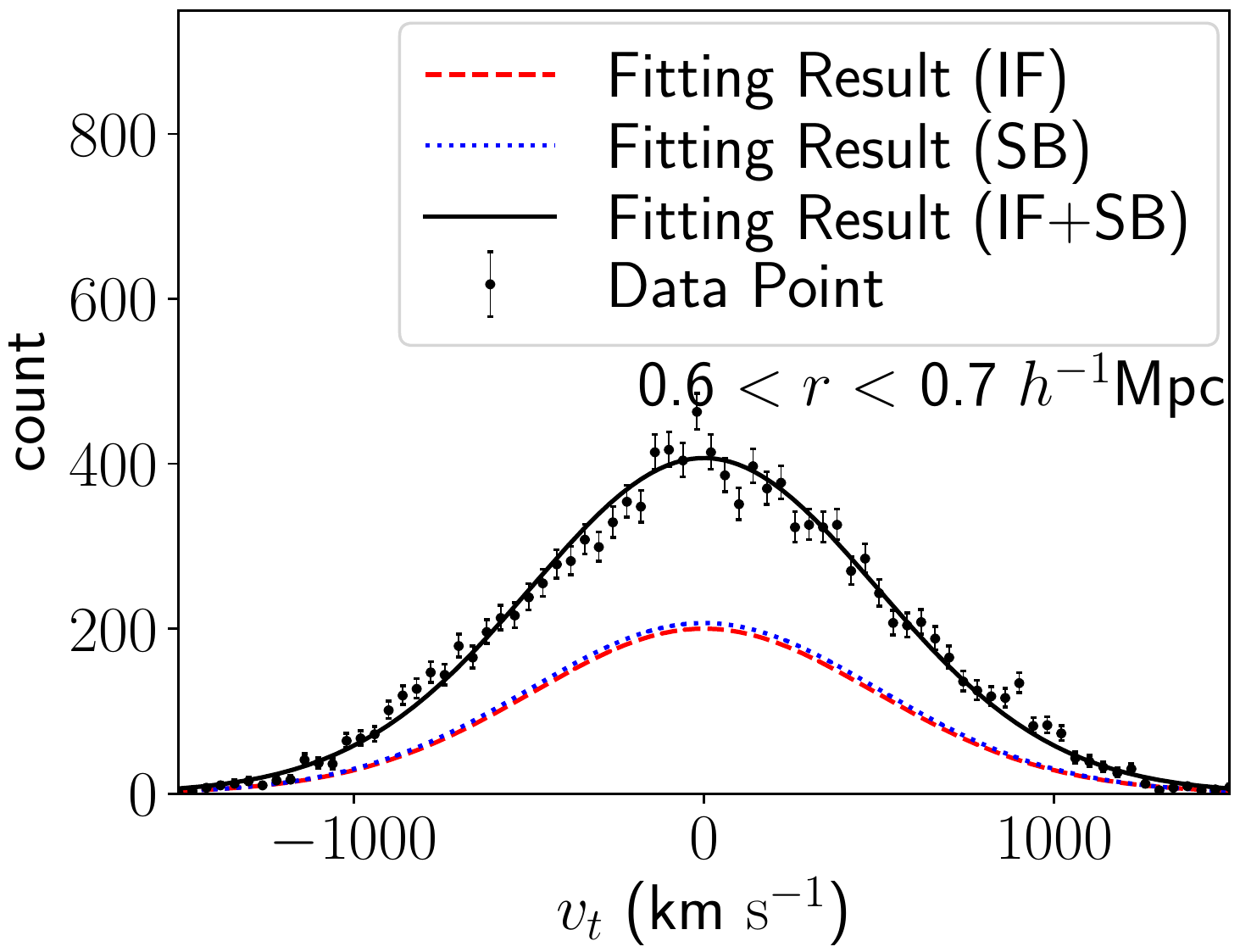}
       \end{minipage}
       \caption{Same as Fig. \ref{FigS-2}, but for the tangential velocity distributions.}

    \label{FigS-3}
\end{figure}

Fig. \ref{FigS-3} shows examples of tangential velocity distributions.
Again, our model function for the tangential velocity distribution is in good agreement with the histogram from our $N$-body simulations.
We interpolate the parameters linearly and use them as smooth functions of $r$. 

\section{Comparison between Observational data and the model}
\label{S-Comp}
\subsection{The Radial Distribution of Red-Spiral Galaxies}
\label{S-CompN}
To complete the model of PDF of $v_{\rm{los}}$ shown in equation (\ref{eqS-01}), we need the radial distribution of red-spiral galaxies.
Assuming the spherical symmetry, we can derive the two-dimensional projected radial distribution of galaxies from the three-dimensional radial distribution as
\begin{equation}
 \label{ObsC-1}
 \Sigma ( r_{\perp} ) =  2 \int ^{ \infty }_{ r_{\perp} } \rho(r) \frac{r}{\sqrt{ r^2 - r_{\perp}^2 }} dr \ ,
\end{equation}
where $\Sigma ( r_{\perp} )$ is the projected surface density of galaxies and $\rho(r)$ is the three-dimensional radial distribution.
We obtain $\rho(r)$ as follows.
First, we assume a model function of $\rho(r)$ of red-spiral galaxies that contains some free parameters.
We then calculate the model $\Sigma ( r_{\perp} )$ by substituting the model function of $\rho(r)$ with free parameters in equation (\ref{ObsC-1}).
Finally, we fit the observed $\Sigma ( r_{\perp} )$ of the red-spiral galaxies with the model $\Sigma ( r_{\perp} )$ to obtain the best fit parameters used in our model $\rho(r)$.
Even though we can reconstruct $\rho(r)$ from the observed $\Sigma ( r_{\perp} )$ in an analytic way by using the Abel integral transform \citep{GalaDy}, we do not adopt this method because our observed $\Sigma ( r_{\perp} )$ of red-spiral galaxies is noisy due to the small number of \tnrv{red-spiral} galaxies.

We adopt three models for $\rho(r)$ of red-spiral galaxies.
The first model is the shell model, whose model function is described as
\begin{equation}
 \label{ObsC-2}
  \rho_{\rm{shell}} (r) = \begin{cases}
  \frac{ C_{\rm{shell}}  \  r^{P_{0}} } { 1 +  \{ ( r - P_{1}) / P_{2} \}^{2} }  & ( r < 2 ~ h^{-1} {\rm{Mpc}}) \\
   A_{0} / ( r + A_{1})  & (  r \geq  2 ~ h^{-1} {\rm{Mpc}}) \ ,
  \end{cases}
\end{equation}
where
\begin{equation}
 \label{ObsC-3}
 C_{\rm{shell}} \equiv \left[ \frac{1 +  \{ ( 2 - P_{1}) / P_{2} \}^{2}}{2^{P_{0}}} \right]   \{ A_{0} / (2 + A_{1}) \}  \ .
\end{equation}
The second one is the cusp model.
We adopt the double power-law density distribution as the distribution of the inner part,
\begin{equation}
 \label{ObsC-4}
  \rho_{\rm{cusp}} (r) = \begin{cases}
  \frac{ C_{\rm{cusp}}  } { (r)^{Q_{0}} \{ 1 +   (r  / Q_{1} ) \}^{Q_{2}} }  & ( r < 2 ~ h^{-1} {\rm{Mpc}}) \\
   A_{0} / ( r + A_{1})  & (  r \geq  2 ~ h^{-1} {\rm{Mpc}}) \ ,
  \end{cases}
\end{equation}
where
\begin{equation}
 \label{ObsC-5}
 C_{\rm{cusp}} \equiv \left[ (2)^{Q_{0}} \{ 1 +   (2  / Q_{1} ) \}^{Q_{2}} \right]  \{ A_{0} / (2 + A_{1}) \}  \ .
\end{equation}
The last one is the core model, whose model function is described as
\begin{equation}
 \label{ObsC-6}
  \rho_{\rm{core}} (r) = \begin{cases}
  \ C_{\rm{core}}  \{ R_{0}  \tanh{ (\frac{ r - R_{1} }{ R_{2} } )} - R_{3} \} & ( r < 2 ~ h^{-1} {\rm{Mpc}}) \\
   A_{0} / ( r + A_{1})  & (  r \geq  2 ~ h^{-1} {\rm{Mpc}}) \ ,
  \end{cases}
\end{equation}
where
\begin{equation}
 \label{ObsC-7}
 C_{\rm{core}} \equiv  \left[ \frac{1} { \{ R_{0}  \tanh{ (\frac{ 2 - R_{1} }{ R_{2} } )} - R_{3} \} } \right]  \{ A_{0} / (2 + A_{1}) \}  \ .
\end{equation}
Note that $A_{0}$, $A_{1}$, $P_{0}$, $P_{1}$, $P_{2}$, $Q_{0}$, $Q_{1}$, $Q_{2}$, $R_{0}$, $R_{1}$, $R_{2}$, and $R_{3}$ are free parameters.
Since we calibrate $A_{0}$ and $A_{1}$ by using only $\Sigma ( r_{\perp} > 2 ~ h^{-1} {\rm{Mpc}}  )$, $A_{0}$ and $A_{1}$ share the same values for all the models.
We adopt these functional forms just to describe $\Sigma ( r_{\perp} )$ of red-spiral galaxies with a small number of free parameters, \tnrv{and are not based on any physical model.}

\begin{figure}
 \begin{minipage}{0.45\textwidth}
    \includegraphics[width=1.0\textwidth]{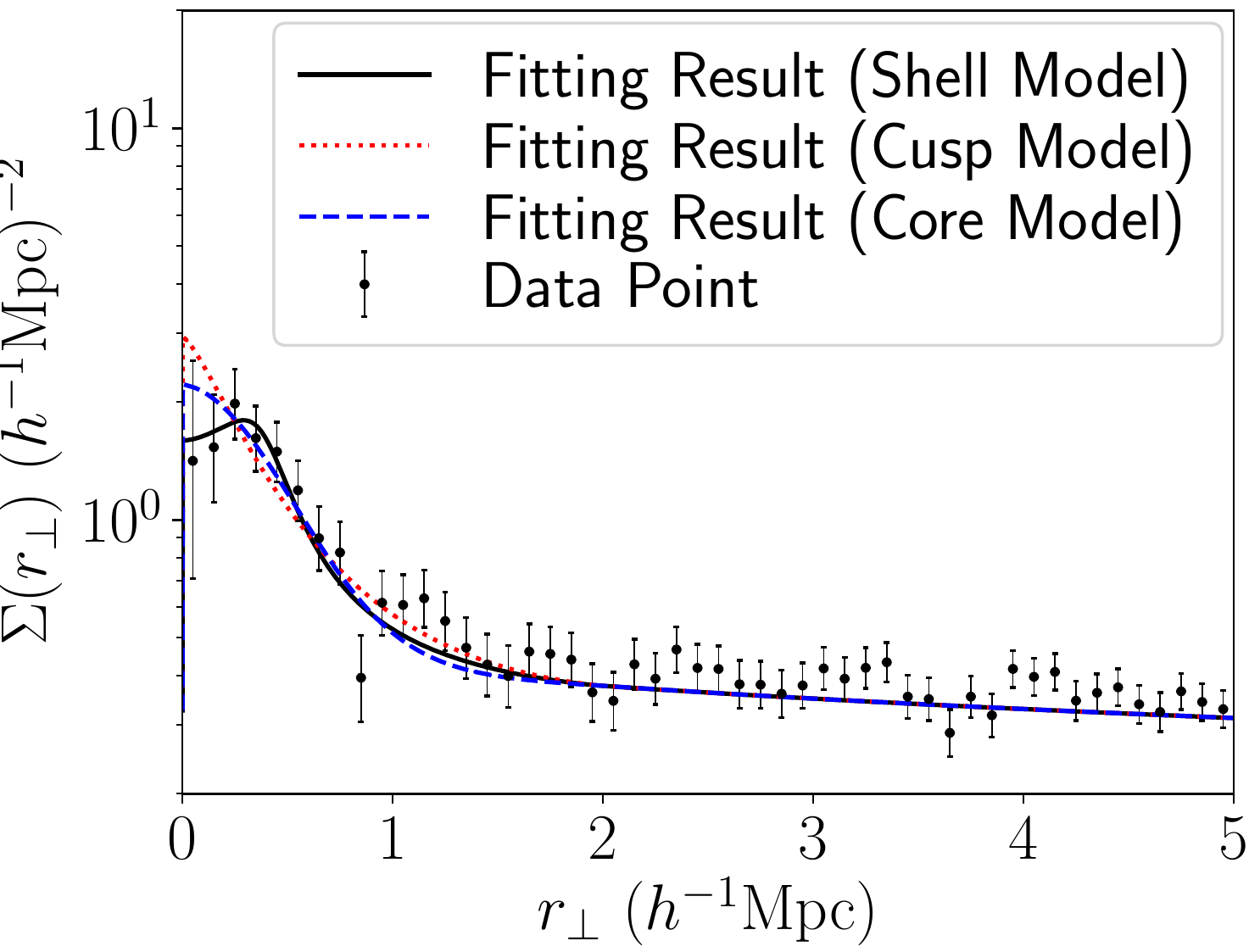}
           \end{minipage}\\
    \begin{minipage}{0.45\textwidth}
    \includegraphics[width=1.0\textwidth]{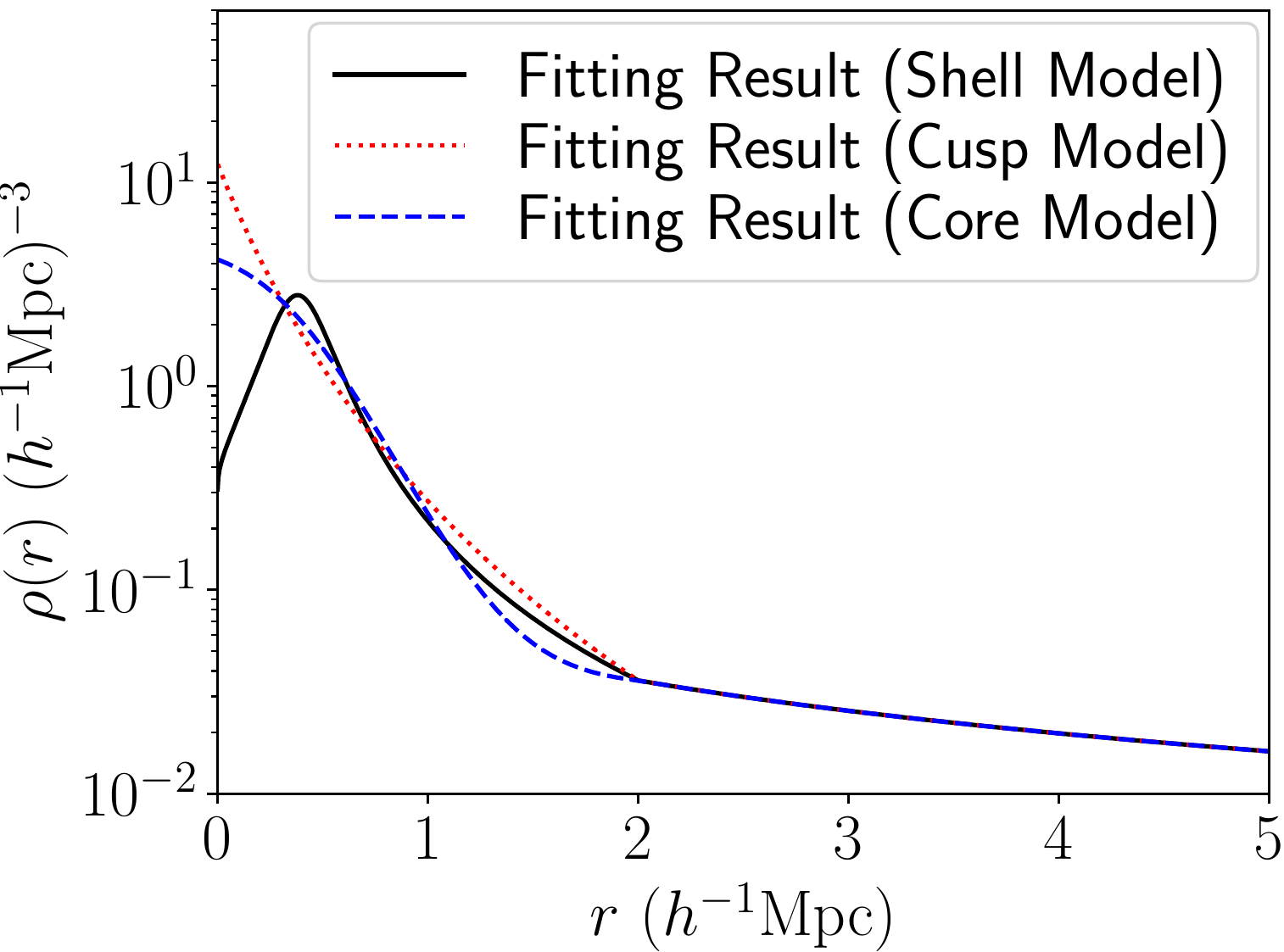}
       \end{minipage}
       \caption{$\it{Top :}$ The projected surface density ($\Sigma ( r_{\perp} )$) of red-spiral galaxies.
Points with error bars show observed $\Sigma ( r_{\perp} )$, the black solid line is the best fit line of the shell model, the red dotted line is the best fit line of the cusp model, and the blue dashed line is the best fit line of the core model.
$\it{Bottom :}$ The radial distribution ($\rho(r) $) of red-spiral galaxies.
The line is the best fit line of each model of $\rho(r)$.}
    \label{FigN-1}
\end{figure}

The top panel of Fig. \ref{FigN-1} shows the observed $\Sigma ( r_{\perp} )$ of red-spiral galaxies.
Again, the error bars are Poisson errors from the number of galaxies in each $r_{\perp}$ bin.
We also show the best fit result of the model $\Sigma ( r_{\perp} )$ for each model.
The degree of freedom for the fitting of the inner part ($0  ~ h^{-1} {\rm{Mpc}} < r < 2.0  ~ h^{-1} {\rm{Mpc}} $) is 17 (16) (16) for the shell (cusp) (core) model, and the $\chi^{2}$ for the inner part is 12.3 (16.7) (14.9) for the shell (cusp) (core) model.
By performing F-test, we find that $\Sigma ( r_{\perp} )$ fitted with the shell model better fits the observed one than the cusp (core) model at the significance of 77.0\% (69.3\%).
This indicates that the shell model is the best model to represent the observed $\Sigma ( r_{\perp} )$ at small $r_{\perp}$, whereas the core and cusp models are also in good agreement with the observed distribution.
The bottom panel of Fig. \ref{FigN-1} shows the model $\rho(r)$ of red-spiral galaxies with best fitting parameters obtained in the top panel of Fig. \ref{FigN-1}.
In the shell model, the radial distribution peaks at $r \sim 0.4  ~ h^{-1} {\rm{Mpc}}  $, and the majority of red-spiral galaxies in the sample are located around the peak.

\subsection{The Phase Space Distribution of Red-Spiral Galaxies}
In Section \ref{S-Sim}, we derive the phase space distribution in three-dimensional space ($p_{v}$) by using cosmological $N$-body simulations.
In Section \ref{S-CompN}, we also derive the radial distributions of red-spiral galaxies ($\rho (r)$) from the observed surface density distribution.
We can calculate the PDFs of $v_{\rm{los}}$ using equation (\ref{eqS-01}) with $p_{v}$ and $\rho (r)$ derived above.

\begin{figure}
 \includegraphics[width=\columnwidth]{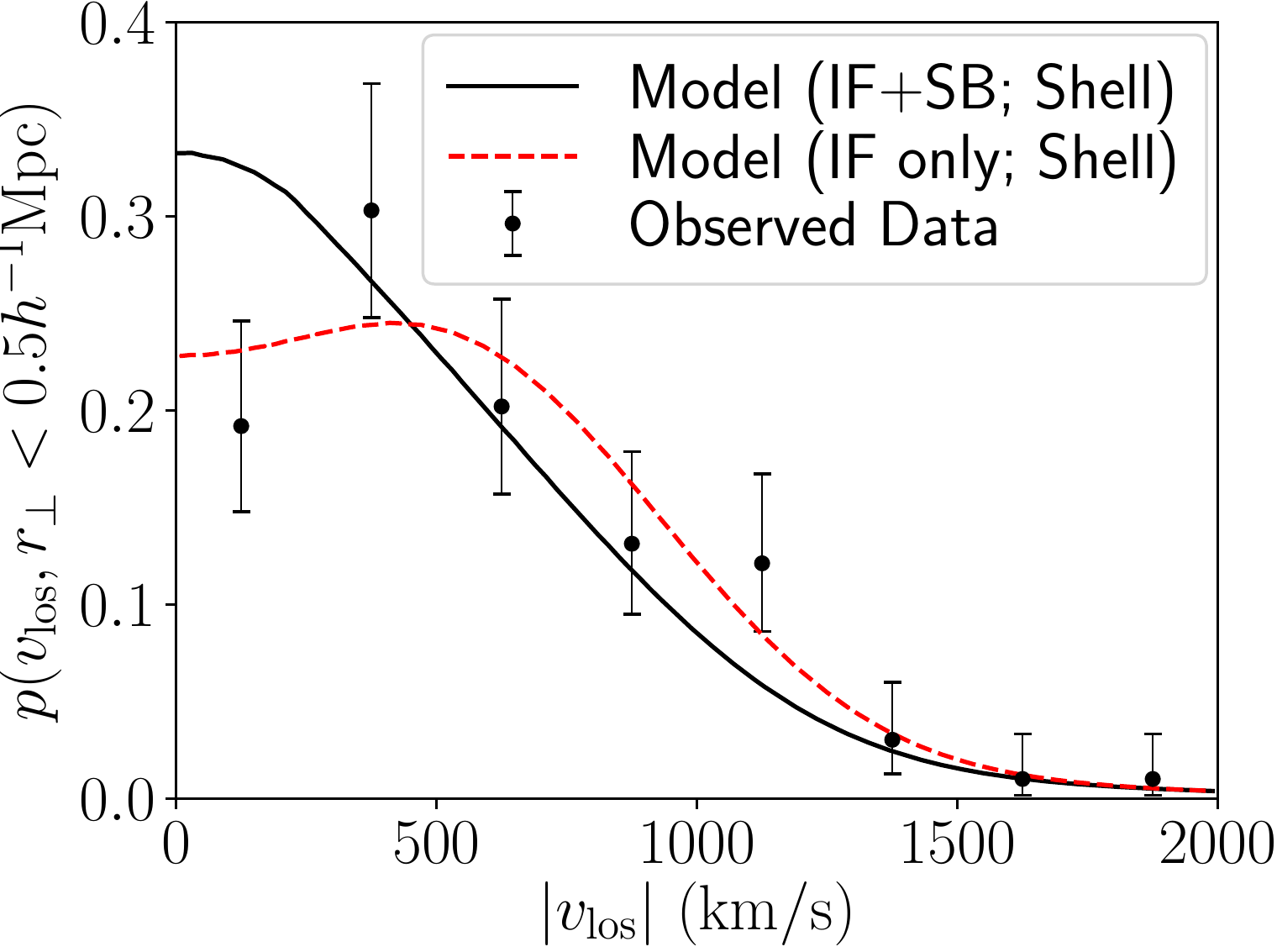}
 \caption{The PDF of $|v_{\rm{los}}|$ of red-spiral galaxies.
Points with error bars are the PDF obtained from the observation, the black solid line is the PDF calculated from equation (\ref{eqS-01}), and the red dashed line is the PDF from the infall component only i.e., PDF calculated from equation (\ref{eqS-01}) with $\alpha (r )= 0$.
We adopt the shell model as the radial distributions of red-spiral galaxies here.
The projected aperture adopted here is $0.5 h^{-1} \rm{Mpc}$.}    
\label{FigPSD-1}
\end{figure}

Fig. \ref{FigPSD-1} shows the PDFs of $|v_{\rm{los}}|$.
The black solid line is the PDF calculated from equation (\ref{eqS-01}) with $\rho_{\rm{shell}} (r)$.
We also show the PDF of $|v_{\rm{los}}|$ calculated from equation (\ref{eqS-01}) with $\alpha (r) = 0$, i.e., the PDF with only the infall component with red dashed line.
This figure indicates that, when we adopt the shell model, the kinematics of infalling galaxies can represent that of red-spiral galaxies, as it nicely reproduces the dip at $v_{\rm los}=0$.
The $\chi^2$ for the PDF by using both the infall and splashback components is 8.8, and that for the PDF by the infall component only is 3.4.
Since the degree of freedom is 7, by performing F-test, we find that the PDF of $|v_{\rm{los}}|$ calculated with only the infall component is preferred over the one with both the infall and splashback components at 89\% significance.
Because the absolute value of the average radial velocity of the infall component is larger than that of the splashback component, the number of galaxies around $|v_{\rm{los}}| = 0$ decreases when we use only the infall component, which helps to reproduce the dip at $v_{\rm los}=0$ as seen in the observed distribution.
\tnrv{Note that the black solid line in Fig. \ref{FigPSD-1} does not correspond to the PDF of $|v_{\rm{los}}|$ for all red galaxies, because we use $\rho (r)$ that reproduces the projected surface density of red-spiral galaxies.}

\begin{figure}
 \includegraphics[width=\columnwidth]{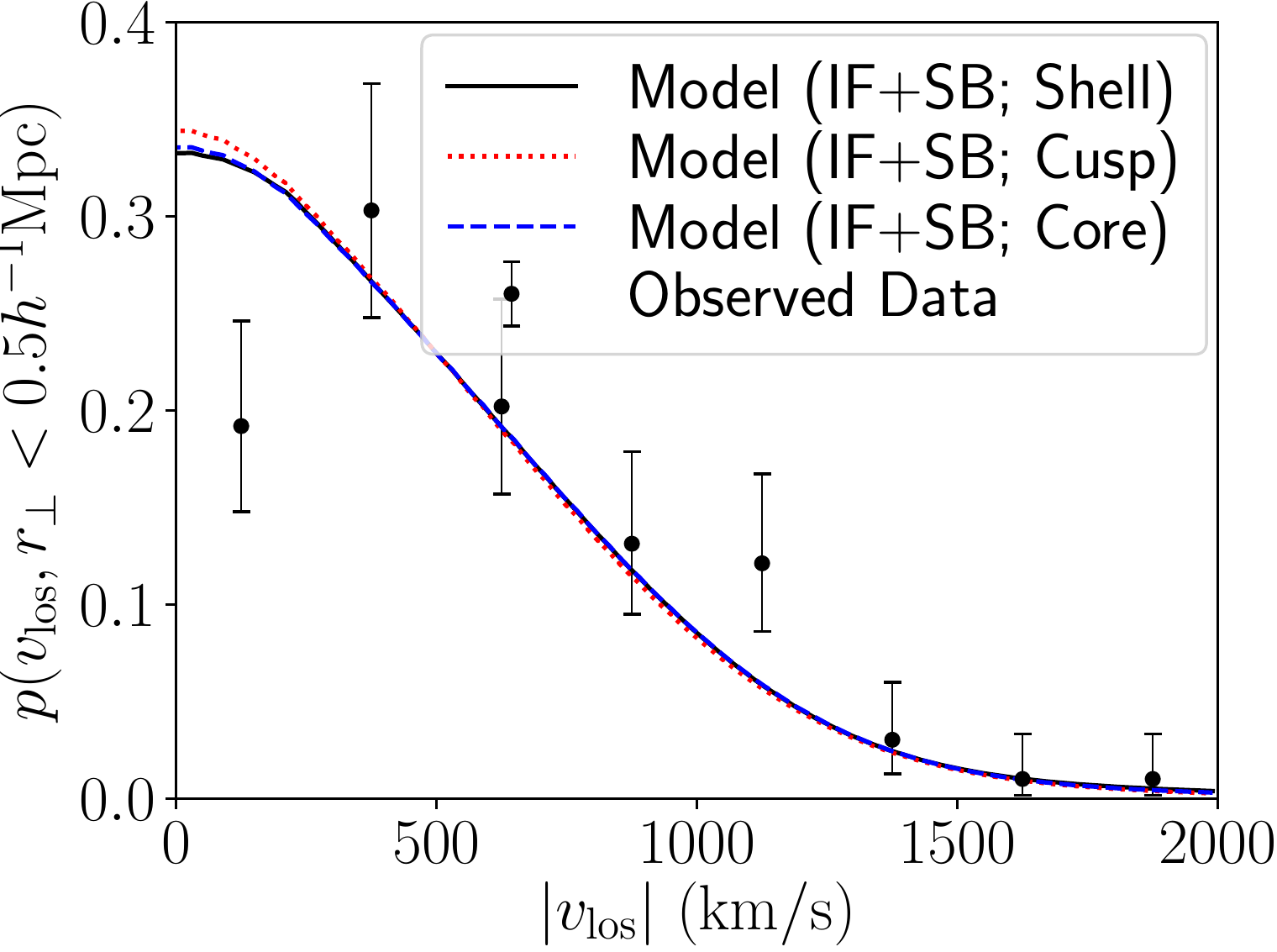}
 \caption{The PDF of $|v_{\rm{los}}|$ of red-spiral galaxies.
Points with error bars are the PDF obtained from the observation, the black solid line is the PDF calculated from equation (\ref{eqS-01}) by using $\rho_{\rm{shell}} (r)$, the red dotted line is the PDF calculated by using $\rho_{\rm{cusp}} (r)$, and the blue dashed line is the PDF calculated by using $\rho_{\rm{core}} (r)$.
Here we use $p_{v}$ with both the infall and splashback components.
The projected aperture adopted here is $0.5 h^{-1} \rm{Mpc}$.}    
\label{FigPSD-2}
\end{figure}

We also check how the choice of $\rho(r)$ affects the PDF of $|v_{\rm{los}}|$.
In Fig. \ref{FigPSD-2}, we compare the PDFs of $|v_{\rm{los}}|$ with each model of $\rho (r)$.
We use $p_{v}$ with both the infall and splashback components here.
This Figure indicates that we cannot reproduce the observed PDF of $|v_{\rm{los}}|$ of red-spiral galaxies regardless of the choice of $\rho (r)$.
The $\chi^2$ for the PDF by using $\rho_{\rm{cusp}} (r)$  is 9.8, and that with $\rho_{\rm{core}} (r)$ is 8.9.

\begin{figure}
 \includegraphics[width=\columnwidth]{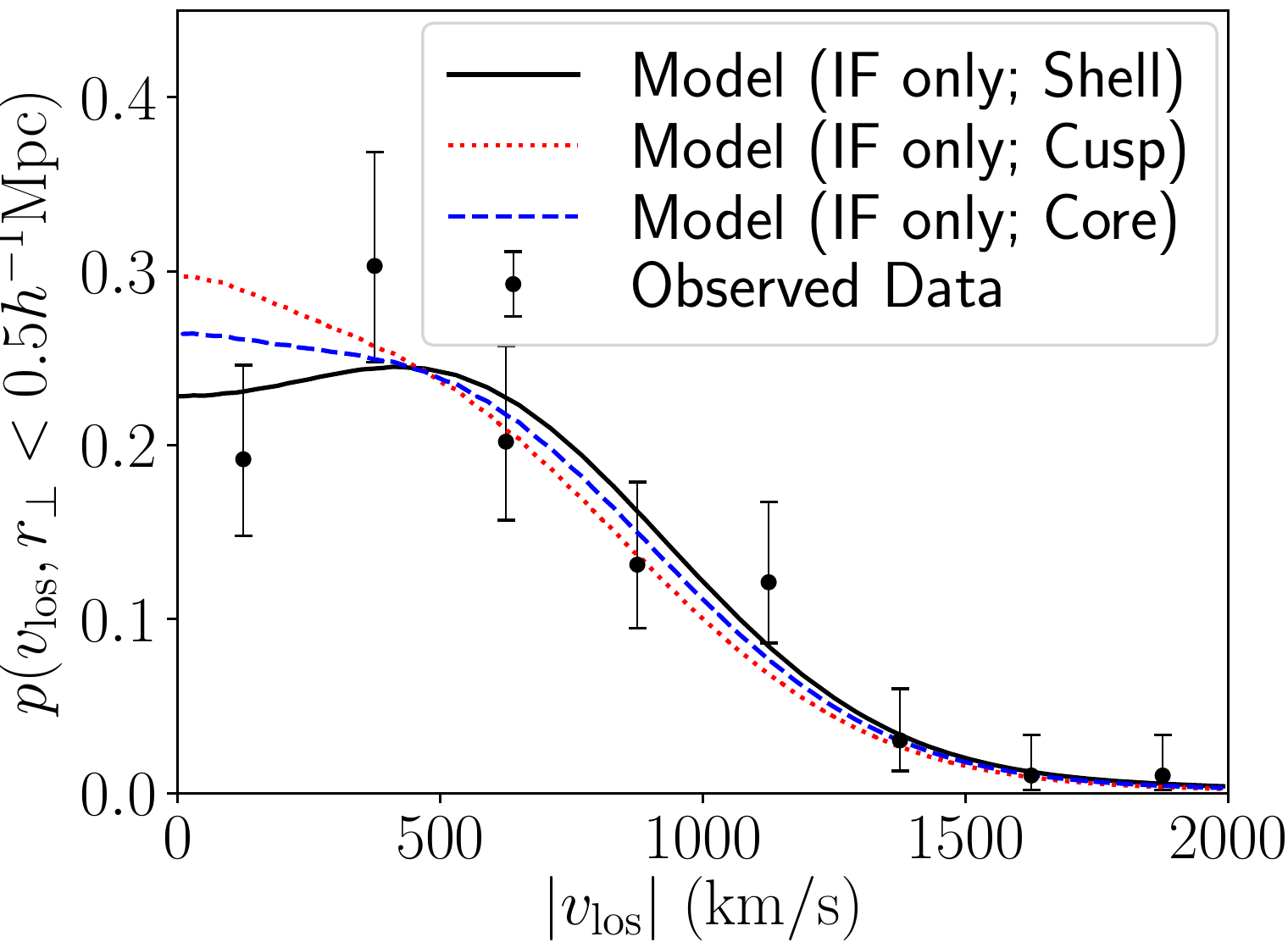}
 \caption{Same as Fig. \ref{FigPSD-2}, but with only the infall component, i.e., $\alpha = 0$.}    
\label{FigPSD-3}
\end{figure}

Fig. \ref{FigPSD-3} shows the comparison of the PDFs of $|v_{\rm{los}}|$ for each model of $\rho (r)$ and only the infall component, i.e., $\alpha = 0$.
We find that only the shell model can reproduce the dip, although the other model can also fit the observed PDF of red-spiral galaxies reasonably well.
The $\chi^2$ for the PDF by using $\rho_{\rm{cusp}} (r)$  is 5.8, and that with $\rho_{\rm{core}} (r)$ is 4.2.
Again, by performing F-test, we find that the PDF of $|v_{\rm{los}}|$ calculated with the shell model is preferred over the cusp (core) model at 76\% (61\%) significance, when we assume that red-spiral galaxies reside only in the infall component.
\tnrv{The different behaviors at small $|v_{\rm los}|$ between different models originate from the different contributions of galaxies located in inner ($r<0.5h^{-1}$Mpc) and outer ($r>0.5h^{-1}$Mpc) regions in the three-dimensional space before the projection.
In the shell model, the contribution of galaxies in the outer region is relatively large, and since radial motions of these outer galaxies are always nearly parallel to the line-of-sight direction they tend to have large $|v_{\rm los}|$, leading the small PDF value at small $|v_{\rm los}|$.}

\section{Discussion}
\label{S-Dis}
\begin{figure}
    \includegraphics[width=\columnwidth]{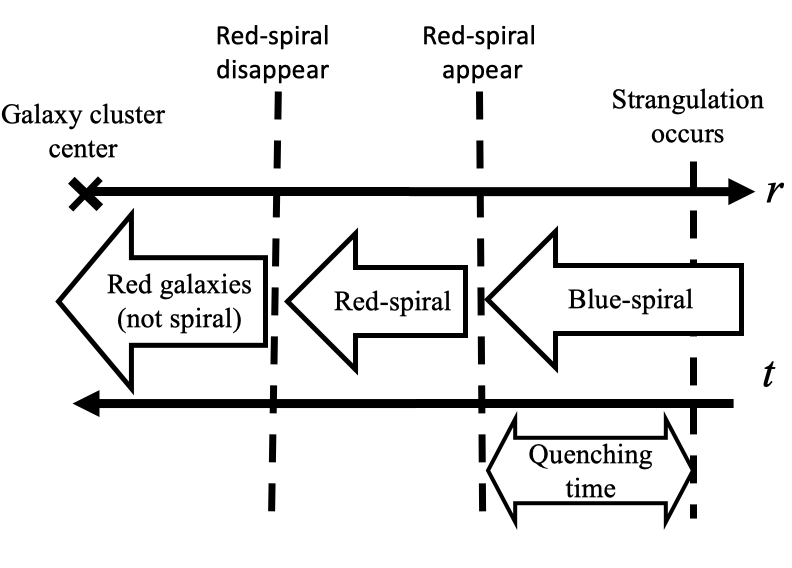}
 \caption{Schematic picture of the scenario to explain red-spiral galaxies.}
 \label{F-Dis-1}
\end{figure}

In this Section, we discuss implications of our results assuming the shell model of the radial distribution of red-spiral galaxies, which is found to best reproduce the observed PDF of $v_{\rm{los}}$.

Our results support the scenario shown in \citet{Bekki2002} that red-spiral galaxies are accreted on to the massive halo as blue-spirals, after which their halo gas are stripped by the strangulation effect without transforming their morphology, and quench the star formation.
When the strangulation effect is the dominant mechanism that turns blue-spiral galaxies into red-spiral galaxies, we can constrain the quenching timescale of galaxies by using the radius where red-spiral galaxies appear, i.e., the larger one of the two half maximum points of the radial distribution of red-spiral galaxies (\tnrv{$r \sim 0.6 h^{-1} \rm{Mpc}$}), and the average value of infalling velocities for each radii obtained from cosmological $N$-body simulations.
Since only hot gas surrounding the galaxies are stripped by the strangulation effect and cold gas remains, blue-spiral galaxies do not turn into red immediately after the strangulation effect occurs.
We regard the quenching time as the time from the strangulation effect occurs to turn blue-spiral galaxies into red (see Fig. \ref{F-Dis-1}).
We can estimate the quenching timescale as \tnrv{$1.0 ~ \rm{Gyr}$ ($2.2 ~ \rm{Gyr}$)} if the strangulation effect occurs at $r = 1 h^{-1} \rm{Mpc}$ ($r = 1.5 h^{-1} \rm{Mpc}$) from the centre of the galaxy cluster.

\tnrv{While our result suggests that the star formation fading process of the satellite is rapid, the timescale on which the quenching of star formation happens is currently heavily debated, and relatively long quenching timescales are proposed.
Theoretical studies with semi-analytic models have revealed that the assumption of instantaneous stripping of the hot halo gas of satellites leads to produce too many quiescent satellite galaxies compared with observations (e.g. \citealt{Weinmann2006}).
Although more gradual hot gas stripping is proposed (e.g. \citealt{Kang2008}; \citealt{Font2008}), these models are still insufficient to reproduce observations.
\citet{DeLucia2012} have compared the semi-analytic galaxy formation models to observational data to predict relatively long quenching timescales of 5-7~Gyr in group and cluster satellites (see also \citealt{Hirschmann2014}).
\citet{Wetzel2013} have constructed an empirical model of the quenching of satellites in clusters, and have shown that star formation rates of satellites evolve unaffected for 2-4~Gyr after infall, after which the star formation rates decrease rapidly.}

\tnrv{On the other hand, recent cosmological hydrodynamical simulations focusing on the quenching in high-density environments suggest relatively short quenching timescales (\citealt{Lotz2018}; \citealt{Arthur2019}).
\citet{Lotz2018} have argued that for almost all satellites, the star formation rate decreases to zero within 1~Gyr after infall due to ram-pressure stripping in their simulation.
\citet{Arthur2019} have shown that most of the satellites in their simulation become gas-poor at ~1.5-2~$r_{200}$ on their first infall, which is consistent with the results of \citet{Lotz2018}.}


\tnrv{Our result indicates that red-spiral galaxies turn into another morphological type of galaxies before they reach the center of galaxy clusters.
Red-elliptical galaxies are candidates of objects after the morphological transformation.
In \citet{Okamoto2001} and \citet{Okamoto2003}, they have suggested that spiral galaxies turn into elliptical galaxies through major mergers.
In addition, several studies (e.g., \citealt{Goto2003}; \citealt{Wolf2009}; \citealt{Masters2010}) have suggested that red-spiral galaxies are progenitors of S0 galaxies.
\citet{Bekki2002} have shown that the spiral arm structure in the red-spiral galaxies disappears due to the decrease of the gas accretion through the strangulation effect, and have suggested that red-spirals gradually evolve into S0s.
Previous observational work has, however, revealed that S0 galaxies have systematically larger bulge fractions, the fraction of the total luminosity of a galaxy associated with the bulge, than spiral galaxies (e.g. \citealt{Christlein2004}).
\citet{Christlein2004} have suggested that it is necessary for spirals to experience the significant bulge growth to be transformed into S0s, and disc-fading models are not enough for a transformation mechanism.
As another mechanism, \citet{Tapia2014} have shown that S0 galaxies are formed through minor mergers.
In addition, galaxy harassment, multiple high-speed encounters (\citealt{Moore1996}; \citealt{Moore1999}), may also be a plausible mechanism transforming red-spirals into S0s.
The galaxy harassment can make discs thicken through tidal heating.
As a result, the spiral structure of the discs disappears, and the morphology of the discs becomes similar to that of S0s. }

\tnrv{While our results do not immediately discriminate these different scenarios of the morphological transformation partly because of the lack of quantitative predictions by these scenarios, we can estimate the radius where the morphological transformation is effective by using our results, which may serve as a useful constraint on models of the morphological transformation.
We regard the radius where the morphological transformation is effective as the radius where red-spiral galaxies disappear, i.e., the smaller one of the two half maximum points of the radial distribution of red-spiral galaxies ($r \sim 0.2 h^{-1} \rm{Mpc}$).}

It is worth noting that since we define spiral galaxies as $p_{\rm{cs \_ debiased}} > 0.8$ based on the Galaxy Zoo, we may miss some spiral galaxies whose spiral arms are hard to detect from images.
This means that the our estimate of the timescale of the morphological transformation might be overestimated.

\section{Summary and Conclusion}
\label{S-Con}
We have performed the stacking analysis of red-spiral galaxies around galaxy clusters by using the Galaxy Zoo and the redMaPPer cluster catalogue.
We have found that the PDF of $v_{\rm{los}}$ of red-spiral galaxies is significantly different from that of all red galaxies, and has a dip at $v_{\rm los}=0$ at $1.4 \sigma$ significance.
To interpret this observation, we have constructed a model of the phase space distribution of galaxies surrounding galaxy clusters in three-dimensional space based on the stacked phase space distribution from cosmological $N$-body simulations.
Following \citet{Hamabata2018}, we have adopted a two component model, which consists of the infall component and the splashback component.
We have investigated the radial distribution of red-spiral galaxies from observed surface density distribution of red-spiral galaxies.
We have considered three model, the shell model, the cusp model, and the core model.
We have found that the PDF of $v_{\rm{los}}$ of red-spiral galaxies can be reproduced by by assuming that red-spiral galaxies reside predominantly in the infall component, particularly for the case of the shell model as it nicely reproduces the central dip of the PDF. 
\tnrv{A caveat is that we have adopted a rather simple model to assign halos in the simulations to red-spiral galaxies.
While this simple model successfully reproduces the observed PDF of $v_{\rm{los}}$ of red-spiral galaxies, it is possible that some other methods to select the haloes of red-spiral galaxies may also reproduce the PDF.}

Our results and analysis suggest that the shell model is the most plausible model of the radial distribution of red-spiral galaxies among the models investigated in this paper, although the core and the cusp models also show reasonably good agreements.
\tnrv{Our results and analysis indicate that the radius where environmental quenching occurs is larger than the radius where the morphological transformation is effective.}
We have found that we can constrain \tnrv{the radius where the quenching occurs as $r \sim 0.6 h^{-1} \rm{Mpc}$}, the quenching timescale by the environmental effect to a few Gyrs, and \tnrv{the radius where the morphological transformation is effective as $r \sim 0.2 h^{-1} \rm{Mpc}$.
This result provides new observational constraints on mechanisms of quenching and morphological transformation.}

We have demonstrated that the detailed analysis of the PDF of $v_{\rm{los}}$ such as the one we have presented in this paper provides much more information on the motion of cluster galaxies than the analysis of just the dispersion of the PDF as often performed in the literature, which provides an important clue to understand the galaxy evolution.
Our results and analysis indicate that the kinematics of galaxies is not fully virialised, but a significant fraction of galaxies are infalling even at $r$ smaller than $0.5 h^{-1} \rm{Mpc}$ at $0.05 < z < 0.1$.
Our results also highlight the fact that we can extract kinematically coherent sample of galaxies by using the features of galaxies such as colours, morphologies, and luminosities, which may help the analysis of the redshift-space distortion effect.
By comparing the observed PDF of $v_{\rm{los}}$ of red-spiral galaxies to that obtained from hydrodynamical cosmological simulations (e.g., \citealt{Schaye2015}, \citealt{Dubois2016}) and cosmological semi-analytic simulations (e.g., \citealt{Henriques2015}, \citealt{Lacey2016}, \citealt{Makiya2016}), we can test the validity of the mechanism of the environmental effect implemented in these simulations.
\tnrv{By increasing the sample of red spiral galaxies, we can obtain the PDF of $v_{\rm{los}}$ of red-spiral galaxies with smaller statistical errors, which will constrain the kinematics and radial distribution of red-spiral galaxies more accurately.
Finally, we may be able to obtain new insights into the origin of S0 galaxies by analyzing phase space distributions of elliptical and S0 galaxies around galaxy clusters in the same way as we did in this paper, which we leave for future work.}

\section*{Acknowledgements}
We thank Taizo Okabe and Shigeki Inoue for useful discussions.
\tnrv{We also thank the anonymous referee for many useful comments and suggestions.}
This work was supported in part by World Premier International Research Center Initiative (WPI Initiative), MEXT, Japan, by MEXT as "Priority Issue on Post-K computer" (Elucidation of the Fundamental Laws and Evolution of the Universe) and JICFuS, and JSPS KAKENHI Grant Number JP15H05892, JP17H02867, JP17K14273, JP18H05437, and JP18K03693.
Numerical simulations were carried out on Cray XC50 at the Center for Computational Astrophysics, National Astronomical Observatory of Japan.


\bibliography{reference}

\bibliographystyle{mnras}



\appendix


\bsp	
\label{lastpage}
\end{document}